\documentclass[aps,prd,preprint,preprintnumbers]{revtex4}
\usepackage{color}
\usepackage{graphicx}

\newcommand{\sgr}{{\rm Sgr~A}^*}
\newcommand{\Msgr}{M_{\rm{Sgr A}^*}}

\begin{document}
\title{Constraining scalar field properties  with boson stars as black hole mimickers}

\author{J. Barranco}
\affiliation{Instituto de Astronom\'{\i}a, 
Universidad Nacional Autonoma de M\'exico, Mexico D.F. 04510, M\'exico}
\author{A. Bernal}
\affiliation{Instituto de Ciencias Nucleares, Universidad Nacional Aut\'onoma de M\'exico, M\'exico D.F. 04510, M\'exico}

\begin{abstract}
Constraints to the mass of a scalar field and the strength of its
self-interacting coupling constant are obtained. 
This was done using observations of stellar dynamics at the center of our galaxy and by
assuming that the dark compact object responsible of such dynamics is a boson star 
and not a supermassive black hole. We show that if such scalar field represents a spin-zero particle with cross section
high enough to be considered collisional dark matter, there is a region of parameters compatible 
with both conditions: that the scalar field play the role of collisional dark matter and that it can form 
objects with the mass and compactness compatible with stellar kinematics. 
\end{abstract}
\maketitle
Particles of zero-spin, represented by a scalar field, have been considered as 
possible dark matter candidates \cite{Boehm, Matos,Briscese:2011ka}.  
This scalar particle could be considered a self-interacting dark matter 
particle (SIDM) \cite{Spergel:1999mh} if its mass $m_\phi$ and its 
self-interacting coupling constant $\lambda$ are such as to produce  an 
elastic cross section as large as  $10^{-23}{\rm cm}^2/{\rm GeV}$.
SIDM is a collisional form of cold dark matter (CDM), originally proposed 
to solve problems that arose when the collisionless CDM theory of structure 
formation was compared with observations of galaxies on small scales. 
In the present note we will construct self-gravitating objects made of 
scalar fields, 
i.e. Boson Stars (BS), 
and by demanding that those BS play the role of two candidates of
super-massive black hole (SMBH), we will constrain $m_\phi$  
and $\lambda$ and compare with the regions needed to be SIDM.

SMBH were primary proposed in order to 
explain the incredible observed amount of energy coming out from
active galactic nuclei.
If SMBH are detectable, it will be only trough their gravitational
effects produced on stars and gas surrounding it or 
trough gravitational wave astronomy. 
In fact, observations on the dynamics of stars and gas close to the center 
of nearby galaxies, give strong evidence of the existence of 
big dark masses, localized in small regions at the center of those 
galaxies and actually it is believed that the same happens in 
most of galaxies.
It is known that dense clusters of compact objects can not 
account for such concentrations of mass because tidal forces would 
disrupt the bodies. This fact leaves as the only possibility that 
the whole mass should correspond to a single object. However, as 
those dark masses are of the order of million of solar masses, they can 
not be a neutron star.
Thus, the argument that these dark objects at the center of galaxies are 
black holes is therefore indirect, and it is based on the elimination 
of other possibilities.
However, we will show here that the current measurements on the 
compactness of those dark compact objects (DCO)
can still be reproduced by self-gravitating objects made of scalar fields. 

Due to the increasing high-angular resolution instrumentation 
developments, it is now possible to study with more detail the innermost
central region of galactic nuclei. In the case of our own galactic center, 
the study of the innermost stellar dynamics has provided strong evidence 
of a DCO associated with the radio source Sagittarius A$^*$ ($\sgr$). 
This is perhaps the strongest dynamical evidence of a SMBH. Based on 
16 years of observation of the so called ``S-stars'', it has been estimated 
the mass of the central DCO of our galaxy to be 
$\sim 4.1 \pm 0.6 \times 10^6 \,
M_{\odot}$~\cite{EisenhauerEtAl05} with radius no larger than  
$6.25$ light-hours~\cite{SchoedelEtAl03}. 

Moreover, the maser emission produced by water molecules in the nucleus of the 
galaxy NGC 4258 delineates a nearly perfect Keplerian thin disk, providing 
the second strongest case for a SMBH candidate. In this case, the mass of 
the central DCO has been estimated to be $M_{\rm{NGC 4258}}=38.1 \pm 0.01 \times
10^6\, M_\odot$, while the observations of the rotation curve 
require a central density of at least $4 \times 10^9\, M_\odot\,
\mbox{pc}^{-3}$, which implies a maximum radius $R_{\rm max}\simeq
36000\,R_{\rm S}$~\cite{Herrnstein:2005xc}.
The other SMBH candidates are typically proved in regions of radius 
$ > 10^5 R_S$, with $R_S$ the Schwarzschild radius. Therefore the evidence 
is less compelling and for this reason we will use only the cases of 
$\sgr$ and the DCO of NGC 4258 to constrain the free parameters of the scalar 
field. 

With that purpose, we will start describing how BS are constructed; later we put constraints to the free parameters of the scalar field using the masses and compactness of $\sgr$ and NGC 4258 and finally we will show that, there is a region in the already restricted space of parameters that allows the scalar field to play the role of SIDM.
\section{Boson stars}
%
Boson stars are stationary solutions to the Einstein equations
$G_{\mu\nu}=8\pi G T_{\mu\nu}$, where $T_{\mu\nu}=\frac{1}{2}[\partial_\mu \Phi^*\partial_\nu\Phi+
\partial_\mu\Phi\partial_\nu\Phi^*]-\frac{1}{2}[\Phi^{*,\alpha}\Phi_{,\alpha}+V(|\Phi|^2)]$ 
is the stress energy tensor of a complex scalar field $\Phi$. 
We will restrict ourselves to the case where the potential of the scalar field is given by  $V(|\Phi|^2)=\frac{1}{2}m_\phi^2|\Phi|^2+\frac{\lambda}{4}|\Phi|^4$, where $m_\phi$ is the mass of the scalar field and $\lambda$ its self-interaction. 
The spherically symmetric case with no self-interaction
($\lambda=0$) was introduced in the late sixties first by Kaup 
\cite{Kaup:1968zz} 
and later studied by Rufinni and Bonazzola \cite{Ruffini:1969qy}. 
Years later the case with self-interaction $\lambda \ne 0$ was considered in 
\cite{Colpi:1986ye}.
The resulting  Einstein equations coupled with the Klein-Gordon equation in 
a spherically symmetric metric $ds^2=-B(r)dt^2+A(r)dr^2+r^2d\Omega^2$, 
and harmonic time dependence of the scalar field 
$\Phi(r,t)=\phi(r) e^{i \omega t}$ are
\begin{eqnarray}\label{sistema}
\frac{A'}{A^2 x}+\frac{1}{x^2}\left(1-\frac{1}{A}\right)-\frac{\sigma'^2}{A}&=&
\left({1 \over \tilde B}+1\right)\sigma^2+{\Lambda \over 2}\sigma^4 \,, \nonumber \\
{\tilde B'\over A\tilde Bx}-{1 \over x^2}\left(1-{1\over A} \right)-\frac{\sigma'^2}{A}&=&
\left({1\over \tilde B}-1\right)\sigma^2-{\Lambda \over 2}\sigma^4 \,, \nonumber \\
\frac{\sigma''}{A}+\left({2\over x}
+{\tilde B'\over 2\tilde B}-{A' \over 2A}\right)\frac{\sigma'}{A}&=&
\left(1-{1\over \tilde B}\right)\sigma+\Lambda\sigma^3\,,
\end{eqnarray}
where the prime denotes derivative with respect to the new variable $x$ 
defined as $x=rm$. 
In the system of equations  (\ref{sistema}) we have defined 
$\sigma=\sqrt{4\pi G}\phi=\sqrt{4\pi m_p^{-2}}\phi$ 
since $G=m_p^{-2}$, with $m_p$ the Planck mass, in the units $\hbar=c=1$. 
For convenience we defined the dimensionless self-interacting coupling 
$\Lambda={\lambda m_p^2 /( 4 \pi m_\phi^2)}$ 
and $\tilde B = m_\phi B/\omega$.
\begin{figure}\label{equilibrio}
\includegraphics[angle=0,width=0.75\textwidth]{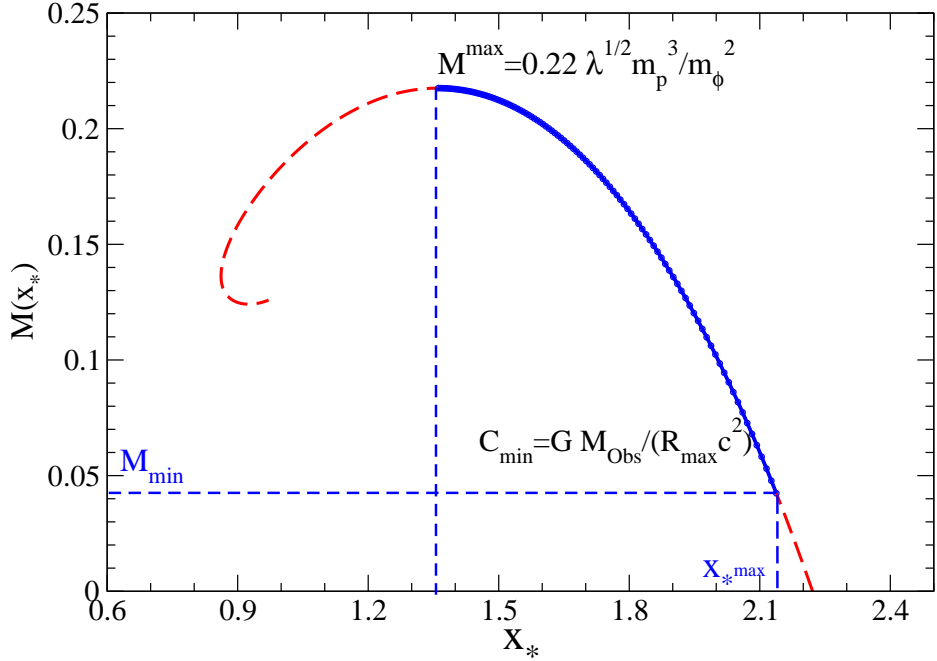}
\caption{Equilibrium configurations for BS in the limit $\Lambda \gg 1$. 
Once the compactness of a DCO is determined, if such compactness is lower
that the maximum compactness of a BS, then it is always possible to find values
$m_\phi$ and $\lambda$ to match the physical mass of the astrophysical object. 
All configurations between 
$M^{\rm min}$ and $M^{\rm max}$ mimic the DCO in consideration.}
\end{figure}
In this work we are interested in constraining scalar field dark matter candidates. It is expected that such scalar fields have very small masses 
compared to $m_p$ and for naturalness $\lambda$ should be of order unity. 
Then $\Lambda$ is expected to be very big. Hence our analysis here will 
concentrate in the case when $\Lambda \to \infty$. See 
\cite{AmaroSeoane:2010qx} for details of the case $\Lambda \sim 1$. 

In the regime \mbox{$\Lambda \gg 1$} it is useful to introduce the new
dimensionless variables~\cite{Colpi:1986ye}
\begin{equation}
\label{reparameter}
\sigma_*=\Lambda^{1/2}\sigma \,, \quad x_*=\Lambda^{-1/2}x\,.
\end{equation}
Rewriting system (\ref{sistema}) in these variables and neglecting terms of 
order $O(\Lambda^{-1})$, the Klein-Gordon
equation can be solved algebraically to yield~\cite{Colpi:1986ye}

\begin{equation}\label{sigmanew}
\sigma_*=\left(\frac{1}{{\tilde B}}-1\right)^{1/2}\,.
\end{equation}
The system (\ref{sistema}) turns out a system of two linear ODE for $A$ and 
${\tilde B}$. Equilibrium configurations are obtained imposing $A(x_*=0)=1$
and an arbitrary value for ${\tilde B(x_*=0)}$. This arbitrariness reflects 
the fact that one can choose freely the central density of the star. 
The full set of configurations are shown in Fig. \ref{equilibrio} where 
$M$ is the mass of the star and $X_*$ is its radius. $M$ is calculated as 
$M=X_*(1-A^{-1}(X_*))/2$ and $X_*$ is defined by the condition 
$\sigma_*(X_*)=0$. Note that the maximum mass allowed for a BS in the limit 
$\Lambda \gg 1$ is $M^{\rm max}\simeq 0.22$.
Finally,  the physical units can be recovered by using the following relations:
\begin{equation}
\label{unitsbiglambda}
\hat{M}=3.24\times10^{51}M\sqrt{\frac{\lambda}{4\pi}}
\left(\frac{\rm eV}{m_\phi}\right)^2 \rm grms\,, \quad
\hat{R}= 2.41 \times 10^{23}X_* \sqrt{\frac{\lambda}{4\pi}}
\left(\frac{\rm eV}{m_\phi}\right)^2 \rm cms\,. 
\end{equation}


Numerical \cite{numerical} and analytical \cite{analytical} 
studies have been done to study the stability of the BS solutions described above.
The result is that configurations with radius larger than the radius that 
corresponds to the maximum mass are stable, see Fig. \ref{equilibrio}, and 
those with smaller radius are not. Only stable BS are used to constrain 
the parameters of the scalar field.  

\section{Limits on $m_\phi$ and $\lambda$}
The limits are obtained as follows:
\begin{itemize}
\item An upper limit is obtained since there is a maximum mass for BS. 
Relations (\ref{unitsbiglambda})
and the maximum mass shown in Fig. \ref{equilibrio} implies
\begin{equation}
m_{\phi} \leq \sqrt{\frac{0.22\, m_{_{P}}^3}{M_{\rm Obs}}}\lambda^{1/4}\,.
\end{equation}
where $M_{\rm Obs}$ is the observed mass of the DCO.
\item The lower limit is obtained as follows\\
1. Compute the minimum observed compactness, defined as the ratio of the 
$M_{\rm Obs}$ over the maximum radius $R_{\rm max}$ where such mass is 
concentrated
\begin{equation}
C^{\rm Obs}_{\rm min}=\frac{G M_{\rm Obs}}{R_{\rm max} c^2}=\frac{M(X_{*\rm max})}{X_{*\rm max}}
\end{equation}
This ``observational'' compactness, defines a minimum numerical mass
$M(X_{*\rm max})$. \\ 
2. This minimum mass defines the lower limit for the scalar field
\begin{equation}
m_\phi \ge \sqrt{\frac{M(X_{*\rm max})m_{_{P}}^3}{M_{\rm Obs}}}\lambda^{1/4}\,.
\end{equation}
\end{itemize}
By applying this simple algorithm to the two most promising candidates 
for SMBH, 
we get the following constrains for $\Msgr$ as:
\begin{equation}
\label{large_lambda_SgrA_max}
3.7 
\times 10^4 \,\lambda^{1/4}~\mbox{eV}
\leq m_{\phi} \leq 
2.9\times 10^5 \,\lambda^{1/4}~\mbox{eV}\,,
\end{equation}
The restriction for NGC 4258, that has as minimum compactness $C_{\rm min}=1/72000$ is :
\begin{equation}\label{large_lambda_NGC}
6.3~ \lambda^{1/4} \mbox{eV} 
\leq m_\phi \leq 
9.6\times 10^4~ \lambda^{1/4}\mbox{eV} \,.
\end{equation}
Finally, by demanding that the scalar field has a cross section 

$\sigma_{2 \to 2}/m_\phi=\lambda^2/(64 \pi m_\phi^2)$,
to be $\sigma_{2 \to 2}/m_{DM}=10^{-25}-10^{-23}~\mbox{cm}^2/\mbox{GeV}$
as required to be considered as collisional dark matter \cite{Spergel:1999mh}, 
it is needed that
\begin{equation}\label{colisional}
9.5 \times 10^{5}\lambda^{2/3} ~{\rm eV}
\leq {m_\phi}
\leq 9.5 \times 10^7 \lambda^{2/3} ~{\rm eV} \,.
\end{equation}

\section{Conclusions}
Combining eqs. (\ref{large_lambda_SgrA_max}-\ref{colisional}), 
there is an intersecting region: $ 6.5 \times 10^{-9} \leq \lambda \leq 4.2 \times 10^{-3}$ and
$332~\mbox{eV} \leq m_\phi \leq 2.46 \times 10^4 ~\mbox{eV}$.
Then, we can conclude that if there exists a scalar field with $m_\phi$ and 
$\lambda$ in this small intersecting region, 
it can form BS as massive and compact as the two best supported SMBH 
candidates and at the same time it can be considered as collisional 
dark matter. 
Nevertheless the region is not compatible with the values found by adjusting
rotation curves in \cite{Arbey:2003sj} and still we are lacking a theory on how these BS are formed in a cosmological context.  

{\bf Acknowledgments:} This work has been supported by CONACyT and SNI-Mexico.


\begin{thebibliography}{99}
\bibitem{Boehm}
  C.~Boehm, T.~A.~Ensslin and J.~Silk,
  J.\ Phys.\ G {\bf 30}, 279 (2004)
  [arXiv:astro-ph/0208458];
  C.~Boehm and P.~Fayet,
  Nucl.\ Phys.\  B {\bf 683}, 219 (2004)
  [arXiv:hep-ph/0305261];
  D.~Hooper, F.~Ferrer, C.~Boehm, J.~Silk, J.~Paul, N.~W.~Evans and M.~Casse,
  Phys.\ Rev.\ Lett.\  {\bf 93}, 161302 (2004)
  [arXiv:astro-ph/0311150];
  C.~Boehm, D.~Hooper, J.~Silk, M.~Casse and J.~Paul,
  Phys.\ Rev.\ Lett.\  {\bf 92}, 101301 (2004)
  [arXiv:astro-ph/0309686];
  C.~Boehm, P.~Fayet and J.~Silk,
  Phys.\ Rev.\  D {\bf 69}, 101302 (2004)
  [arXiv:hep-ph/0311143].

\bibitem{Matos}
T.~Matos, L.~A.~Urena-Lopez,
Class.\ Quant.\ Grav.\  {\bf 17}, L75-L81 (2000). [astro-ph/0004332];
T.~Matos, L.~A.~Urena-Lopez,
Phys.\ Rev.\  {\bf D63}, 063506 (2001).
[astro-ph/0006024];
A.~{Bernal}, T.~{Matos}, and D.~{N{\'u}{\~n}ez},
Revista Mexicana de Astronom{\'\i}a y Astrof{\'\i}sica {\bf 44}, 149 (2008).


\bibitem{Briscese:2011ka}
  F.~Briscese,
  Phys.\ Lett.\  {\bf B696}, 315-320 (2011).
  [arXiv:1101.0028 [astro-ph.CO]].
\bibitem{Spergel:1999mh}
  D.~N.~Spergel, P.~J.~Steinhardt,
  Phys.\ Rev.\ Lett.\  {\bf 84}, 3760-3763 (2000).
  [astro-ph/9909386].


\bibitem{EisenhauerEtAl05} 
 F.~Eisenhauer
{\it et al.},
  Astrophys.\ J.\  {\bf 628}, 246-259 (2005).
  [astro-ph/0502129];
 A.~M.~Ghez
{\it et al.},
  Astrophys.\ J.\  {\bf 689}, 1044-1062 (2008). [arXiv:0808.2870 [astro-ph]];
S.~Gillessen, F.~Eisenhauer, S.~Trippe, T.~Alexander, R.~Genzel, F.~Martins, T.~Ott,
  Astrophys.\ J.\  {\bf 692}, 1075-1109 (2009).
  [arXiv:0810.4674 [astro-ph]];
A.~M. {Ghez}, S.~{Salim}, S.~D. {Hornstein}, A.~{Tanner}, J.~R. {Lu},
  M.~{Morris}, E.~E. {Becklin}, and G.~{Duch{\^ e}ne}, 
  Astrophys. J. {\bf 620}, 744-757 (2005).

\bibitem{SchoedelEtAl03}
R.~{Sch{\"o}del}, T.~{Ott}, R.~{Genzel}, A.~{Eckart}, N.~{Mouawad}, and
  T.~{Alexander}, 
Astrophys. J. {\bf 596}, 1015-1034 (2003);
A.~M.~Ghez, G.~Duchene, K.~Matthews, S.~D.~Hornstein, A.~Tanner, J.~Larkin, M.~Morris, E.~E.~Becklin {\it et al.},
  Astrophys.\ J.\  {\bf 586}, L127-L131 (2003).
  [astro-ph/0302299].

\bibitem{Herrnstein:2005xc}
J.~R.~Herrnstein, J.~M.~Moran, L.~J.~Greenhill, A.~S.~Trotter,
  Astrophys.\ J.\  {\bf 629}, 719-738 (2005).
  [astro-ph/0504405].

\bibitem{Kaup:1968zz} 
D.J. Kaup,  Phys.\ Rev.\  {\bf 172}, 1331 (1968); 

\bibitem{Ruffini:1969qy}
  R.~Ruffini and S.~Bonazzola,
  Phys.\ Rev.\  {\bf 187}, 1767 (1969).

\bibitem{AmaroSeoane:2010qx}
  P.~Amaro-Seoane, J.~Barranco, A.~Bernal and L.~Rezzolla,
  JCAP {\bf 1011}, 002 (2010)
  [arXiv:1009.0019 [astro-ph.CO]].

\bibitem{Colpi:1986ye}
  M.~Colpi, S.~L.~Shapiro and I.~Wasserman,
  Phys.\ Rev.\ Lett.\  {\bf 57} (1986) 2485.

\bibitem{numerical}
  E.~Seidel and W.~M.~Suen,
  Phys.\ Rev.\  D {\bf 42}, 384 (1990).
  S.~H.~Hawley and M.~W.~Choptuik,
  Phys.\ Rev.\  D {\bf 62}, 104024 (2000)
  [arXiv:gr-qc/0007039].

\bibitem{analytical}
  M.~Gleiser and R.~Watkins,
  Nucl.\ Phys.\  B {\bf 319}, 733 (1989).
  T.~D.~Lee and Y.~Pang,
  Nucl.\ Phys.\  B {\bf 315}, 477 (1989).
  F.~V.~Kusmartsev, E.~W.~Mielke and F.~E.~Schunck,
  Phys.\ Rev.\  D {\bf 43}, 3895 (1991)
  [arXiv:0810.0696 [astro-ph]].

\bibitem{Arbey:2003sj}
  A.~Arbey, J.~Lesgourgues, P.~Salati,
  Phys.\ Rev.\  {\bf D68}, 023511 (2003).
  [astro-ph/0301533].
\end{thebibliography}
\end{document}